\documentclass[letterpaper, 11 pt]{article}                                                          

\usepackage{graphicx}                                                          
\usepackage{amsmath,amssymb}
\usepackage[noadjust]{cite}
\usepackage{subfigure}                                       
\usepackage{xy}
\usepackage{array}
\usepackage{fancyhdr}
\usepackage{html,makeidx}                                                      

\date{}

\title{\Huge \bf
Computer Algebra Methods in Control Systems}

\date{ }

\author{Masoud Abbaszadeh
\thanks{M. Abbaszadeh is with GE Global Research, NY, USA, {\tt\small e-mail: masoud@ualberta.net}}}

\begin{document}

\maketitle \thispagestyle{empty} \pagestyle{empty}

\begin{abstract}
As dynamic and control systems become more complex, relying purely on numerical computations for systems analysis and design might become extremely expensive or totally infeasible. Computer algebra can act as an enabler for analysis and design of such complex systems. It also provides means for characterization of all solutions and studying them before realizing a particular solution. This note provides a brief survey on some of the applications of symbolic computations in control systems analysis and design.
\end{abstract}

\section{Introduction}

Computer Algebra is a branch of computer science that deals with symbolic computations. In symbolic computations, mathematical expressions are stored, manipulated and verified using symbolic variables, as opposed to the numerical variables in numerical computations. There is a large potential for utilizing symbolic computations in different areas of control systems engineering especially in nonlinear control and parametric robust control. This has gained research attraction both in controls and advanced computing communities. From theoretical point of view, symbolic methods contribute to control systems in two frameworks: a) a symbolic approach is taken to (eventually) produce a numerical result which could be extremely difficult (if not impossible) to be found via purely numerical methods, and b) a symbolic approach is taken to produce a symbolic result.

The increasing complexity associated with many modern
engineering applications, including autonomous robot
guidance and navigation, self-driving cars, process control in sensor-rich environments,
cyber-physical security of industrial control systems, smart power grid, and control of biological systems, has far-reaching
implications for control system design. As an example, reactive,
embedded software systems, interacting among themselves and
remote users over communication networks, introduce a whole
new set of system-level challenges, and classic control design
objectives such as stability, performance, and robustness are
being complemented with a number of new questions. These
include the cost of hardware implementation, measured for
example not only by computational requirements such as speed
and memory, but also by communication requirements such as
available communication bandwidth. Moreover, the complexity
associated with specifying the control procedures and with
verifying the behavior of the closed-loop system increasingly
plays a fundamental role, especially in safety-critical control
systems arising in energy and transportation networks, and in
medical applications \cite{IEEESymbolic}.

During the last decade, significant progress has been made
towards addressing these issues and overcoming the complexity
associated with such novel control tasks. This complexity stems
from a number of sources, including the complexity of the task
itself, the complexity of the system dynamics, and the complexity
of the environment in which the system is deployed. An
emerging approach to address the complexity issue is to decompose
the control task into a finite collection of building blocks, or
modes of operation. As a result, control procedures are no longer
solely thought of as mappings from sensory data to actuator
signals, but rather as sequences of tokenized instructions that
contain descriptions of such mappings \cite{IEEESymbolic}.

Furthermore, in the age of \emph{Big Data} and connectivity, control systems connected to an industrial internet of things (IIoT), utilize and analyse massive amounts of data in real-time. This opens opportunities for adaptive and online learning of the systems behaviour through real-time system identification via techniques such as reinforcement learning. To provide the right level of abstraction required for these tasks and overcome the curse of dimensionality, it is often desired to develop parameterized models for the plant or controller model and evaluate/adapt those parametric forms in real-time.

The general area of ``Symbolic Control'' has been developed
under the banner of ``Hybrid Systems''. Such systems are
systems that are influenced and characterized by models with
both discrete and continuous components, from switched linear
systems to full-scale hybrid automata. A number of results
have emerged in this area with a classic control-theoretic flavor,
including optimal control, stability, system identification,
observers, and well-posedness of solutions. However, a new
line of research in hybrid systems has also been launched that
studies issues not quite standard to the controls community,
including formal verification, abstractions, model expressiveness,
computational tools, and specification languages. These
latter results belong to the class of results that we refer to as
symbolic control, and what makes them different from the first
class of problems is that they address questions at the highest
level, i.e., at the level of symbols, and as such draw on tools
from computer science and discrete mathematics as much as on
classic control theory. At the same time, they provide faithful
descriptions of the continuous level performance of the actual
system, and as such, provide a formal bridge between the
continuous and the discrete \cite{IEEESymbolic}.

\section{Advantages of the Symbolic Approach}

Symbolic and hybrid symbolic-numeric methods can be vastly superior to the purely numerical methods
in various aspects. The symbolic approach's help is two-fold i) characterizing all possible solutions for a
specific control problem by providing an analytic solution ii) empowering parametric analysis and design.
Depending on the specific method used, the advantage of the parametric design ca be, in turn, two-fold:

 \begin{itemize}
   \item Analysis and design when the system model has symbolic parameters (e.g. unknown, uncertain or time varying parameters in the plant)
   \item Analysis and design when the design specifications (the desired criteria that must be satisfied by the controller) are parametric
 \end{itemize}

An example of the latter is the symbolic pole placement using state feedback when the desired
locations of the closed-loop poles are given based on symbolic parameters.

Symbolic methods are normally used to compute symbolic/parametric results. However, they can also be utilized to efficiently compute numerical results
in cases that a fully numerical approach is very tedious if not impossible. This is specifically true when fixed-structure low order controllers are to be designed to control higher order plants with comprehensive controller stability and performance requirements. For instance, PID tuning for pole placement in a desired region specified by natural frequency and damping coefficients, is an extremely difficult numerical problem which demands vast computational power and large number of simulations \cite{Bhattacharyya}. Nonetheless, the problem can be efficiently solved using parametric polynomial root finding (based on cylindrical algebraic decomposition) to not only find desired values for the PID gains but also characterize the feasible region of the PID gains for which the design specifications (in this case, the close loop poles being in the specified desired region) are satisfied. Therefore, the symbolic approach can significantly enhance control design practices in
which the final design is often obtained through an iterative design and simulation process.
The following are some of the application areas of computer algebra in control theory to mention a few:

\begin{itemize}
  \item Symbolic global optimization methods for optimal control
  \item Symbolic modeling and analysis of nonlinear systems e.g.
        Lyapunov stability analysis and calculation of the region of attraction
  \item Robust control under parametric uncertainty
  \item Pole placement design in an \emph{arbitrary desired region} of the L.H.P based on symbolic
        root finding of polynomial systems
  \item Symbolic calculation of Jacobian for analytical linearization of nonlinear systems
  \item Symbolic calculation in geometric nonlinear control
  \item Symbolic calculation of Jacobian and Hessian matrices for real-time optimization such as in model predictive control (MPC).
\end{itemize}

\section{Illustrations of Symbolic Methods in Control}

Some of the main areas of the symbolic methods in control analysis and design are briefly discussed in this section.

\subsection{Nonlinear Control}

A straightforward but very important of using symbolic computation in nonlinear control is
the analytic calculation of the Jacobian matrix. A well-known presentence of the Jacobian matrix is in the linearization of nonlinear dynamical systems. However, it has applications in the direct nonlinear control and observer design methods, as well.

An important class of nonlinear systems of great theoretical and practical importance are the so-called Lipschitz systems which satisfy a Lipschitz or one-sided Lipschitz continuity condition. A large body of literature elitists addressing analysis and control these systems. See for example \cite{Rajamani5,abbaszadeh2008robust,abbaszadeh2007robust,Raghavan,abbaszadeh2006robust,Xu1,Xu2,Xu3,Xu4,Hammouri,Lu,Gao,
abbaszadeh2010nonlinear,abbaszadeh2008lmi,Thau1,abbaszadeh2010dynamical,abbaszadeh2010robust2,Rajamani2,Abbaszadeh_Chapter2012,
deSouza1,deSouza2,abbaszadeh_phdthesis,abbaszadeh2012generalized,Abbaszadeh5}. However, verifying this property on a given real-world system and estimating its associated parameters, still remains a challenging task. Consider the following continuous-time nonlinear dynamical system
\begin{align}
\dot{x}(t)&=Ax(t)+ \Phi(x,u)\hspace{7mm} A \in\mathbb{R}^{n\times
n}\label{con1}\\
y(t)&=Cx(t)\hspace{25mm} C \in\mathbb{R}^{n\times p},\label{con2}
\end{align}
where $x\in {\mathbb R} ^{n} ,u\in {\mathbb R} ^{m} ,y\in {\mathbb
R} ^{p} $ and $\Phi(x,u)$ represents a nonlinear function that is
continuous with respect to both $x$ and $u$. The system (\ref{con1})-\eqref{con2}
is said to be \emph{locally
Lipschitz} in a region $\mathcal{D}$ including the origin with
respect to $x$, uniformly in $u$, if there exist a constant $l>0$ satisfying:
\begin{eqnarray}
\|\Phi(x_{1},u^{*})-\Phi(x_{2},u^{*})\|\leqslant l\|x_{1}-x_{2}\|
\hspace{7mm}\forall \, x_{1} (t),x_{2} (t)\in \mathcal{D},\label{Lip}
\end{eqnarray}
where  $u^{*}$ is any admissible control signal. The smallest constant $l>0$ satisfying (\ref{Lip}) is known as the
\emph{Lipschitz constant}.
The region $\mathcal{D}$ is the \emph{operational region} or our \emph{region of interest}. If the condition
(\ref{Lip}) is valid everywhere in $\mathbb{R}^{n}$, then the function is said to be globally Lipschitz.

The following definition introduces one-sided Lipschitz functions:

The nonlinear function $\Phi(x,u)$ is said to be
\emph{one-sided Lipschitz} if there exist $\rho \in \mathbb{R}$ such that \cite{Hairer2}
\begin{eqnarray}
\left<\Phi(x_{1},u^{*})-\Phi(x_{2},u^{*}), x_{1}-x_{2}\right> \
\leqslant \rho \|x_{1}-x_{2}\|^{2} \hspace{7mm}\forall \, x_{1},x_{2}\in \mathcal{D},\label{con3}
\end{eqnarray}
where $\rho \in \mathbb{R}$ is called the \emph{one-sided Lipschitz constant}. As in the case of Lipschitz functions,
the smallest $\rho$ satisfying \eqref{con3} is called  the one-sided Lipschitz constant.\\

Similarly to the Lipschitz property, the one-sided Lipschitz property might be local or global.
Note that while the Lipschitz constant must be positive,
the one-sided Lipschitz constant can be positive, zero or even negative.
For any function $\Phi(x,u)$, we have:
\begin{align}
|\left<\Phi(x_{1},u^{*})-\Phi(x_{2},u^{*}), x_{1}-x_{2}\right>| &\leqslant \|\Phi(x_{1},u^{*})-\Phi(x_{2},u^{*})\|\|x_{1}-x_{2}\|\notag\\
\text{and if $\Phi(x,u)$ is Lipschitz, then:} \ \ \ \ \ &\leqslant  l \|x_{1}-x_{2}\|^{2}.
\end{align}
Therefore, any Lipschitz function is also one-sided Lipschitz. The converse, however, is not true.
For Lipschitz functions,
\begin{align}
-l \|x_{1}-x_{2}\|^{2} \leqslant \left<\Phi(x_{1},u^{*})-\Phi(x_{2},u^{*}), x_{1}-x_{2}\right> \
\leqslant l \|x_{1}-x_{2}\|^{2},
\end{align}
which is a \emph{two-sided} inequality v.s. the \emph{one-sided} inequality in \eqref{con3}.
If the nonlinear function $\Phi(x,u)$ satisfies the one-sided Lipschitz
continuity condition globally in $\mathbb{R}^{n}$, then the results are valid globally.
For continuously differentiable nonlinear functions it is well-known that the smallest possible
constant satisfying (\ref{Lip}) ({\it i.e.}, the Lipschitz constant) is the supremum of the norm of
Jacobian of the function over the region $\mathcal{D}$ (see for example \cite{Marquez}), that is:
\begin{align}
l= \limsup \left(\left\|\frac{\partial \Phi}{\partial x}\right\|\right), \ \ \ \ \forall x \in \mathcal{D}.
\end{align}
Alternatively, the one-sided Lipschitz constant is associated with the \emph{logarithmic matrix norm (matrix measure)} of the Jacobian.
The logarithmic matrix norm of a matrix $A$ is defined as \cite{Dekker}:
\begin{align}
\mu(A)= \lim_{\epsilon \rightarrow 0} \frac{|||I+\epsilon A |||-1}{\epsilon},\label{mu1}
\end{align}
where the symbol $|||.|||$ represents any matrix norm.
Then, we have \cite{Dekker}
\begin{align}
\rho= \limsup \left[\mu \left(\frac{\partial \Phi}{\partial x}\right)\right], \ \ \ \ \forall x \in \mathcal{D}.
\end{align}
If the norm used in \eqref{mu1} is indeed the induced 2-norm (the spectral norm) then it can be shown that $\mu(A)=\lambda_{max}\left(\frac{A+A^{T}}{2}\right)$ \cite{Vidyasagar1}.
On the other hand, from the Fan's theorem (see for example \cite{Horn1}) we know that for any matrix, $\lambda_{max}\left(\frac{A+A^{T}}{2}\right) \leq \sigma_{max}(A)=\|A\|$ \cite{Horn1}. Therefore $\rho \leq l$. Usually one-sided Lipschitz constant can be found to be much smaller than the Lipschitz constant \cite{Dekker}. Moreover, it is well-known in numerical analysis that for stiff ODE systems, $\rho << l$ \cite{Stuart, Dekker}. The one-sided Lipschitz continuity plays a vital role in solving such initial value problems.

According to the above discussion, calculation of the Jacobian and the norm of the Jacobian plays a vital role in the estimation of the Lipschitz and one-sided Lipschitz constants which in turn are essential for control of such systems. It is worth mentioning that numerical estimation of the these constants is a global optimization problem, often difficult to solve \cite{Wood}. The Lipschitz one-sided Lipschitz constants constants are usually region-based i.e. they are described the based on the operating region and vise versa. Therefore, providing analytic descriptions of this constants through symbolic computations, also provides insight about the region of operation.

Other important applications of the symbolic methods in nonlinear control include symbolic formation of Lyapunov stability candidates, computation of coordinate transformation for feedback linearization and generation of parametric reference trajectories.

\subsection{Robust Control under Parametric Uncertainty}

The $H_{\infty}$ robust control approach based on the numerical solution of the algebraic Riccati equations or the equivalent linear matrix inequalities (LMIs) is now well established. However, in recent years, another approach to robust stability and performance has emerged based on interval polynomials that can directly address real uncertainty in system parameters and can result in much simpler controllers. The familiar Nyquist diagram, Bode plots and Evans root locus have all been generalized, in this framework, to systems where the coefficients of the polynomials describing the elements of the system are uncertain, in the sense that they are not known exactly but are known to lie within certain limits. This situation arises frequently in practice due to our inability to exactly model the dynamics of systems (e.g. due to high cost or complexity), parameter variations as the system moves within its operating region, or can arise due to manufacturing tolerances or performance variations of a component/subsysem \cite{Munro1}.

Symbolic methods have been studied and successfully applied to parametric robust control problems both for fixed structure lower order controller design and generic higher order robust control design mostly based on parametric root finding for interval polynomials \cite{Munro1,Munro2}. More recently, some parametric robust control methods have been also developed using quantifier elimination \cite{Hyodo}.

\subsection{Symbolic Methods in Hybrid Systems Control}

Over the last few years, the interest in the study of dynamical processes of a mixed
continuous and discrete nature, denoted as hybrid systems, have been growing fast both
in academia and in industry. Hybrid systems consist of continuous dynamical systems
such as differential/difference equations and discrete dynamical systems
such as if-then rules and finite automata.
Hybrid systems are characterized
by the interaction of continuous models, describing continuous
variables governed by differential or difference equations, and
of discrete models, describing symbolic variables governed by
logic rules, switching mechanisms, and other discrete behaviors.
Hybrid systems can switch between many operating modes
where each mode is governed by its own characteristic continuous
dynamical laws. Mode transitions may be endogenous
(variables crossing specific thresholds), or exogenous (discrete
commands directly given to the system). The interest in hybrid
systems is mainly motivated by the large variety of practical
situations where physical processes interact with digital controllers,
as for instance in embedded control systems \cite{Bemporad}.

Combinatorial optimization over continuous and integer
variables is a useful tool for solving complex optimal control
problems of hybrid dynamical systems formulated in discrete time.
Hybrid symbolic-numeric optimization techniques have been successfully applied to complex optimal and
model predictive control of hybrid systems and shown to be superior compared to full numeric
linear, quadratic and mixed integer programming techniques \cite{Bemporad, Kobayashi}.

\subsection{Discretization of LTV and LPV Models}
Liner time-varying (LTV) and linear parameter-varying (LPV) models are often used in adaptive and gain-scheduling control. These models are often developed using first-principles or grey-box modeling techniques and hence are naturally in continuous-time domain. However, the real-time controls is implemented and applied in discrete-time domain. The controller utilizing these models is time-varying normally changing in every sampling time. So, depending whether the digital controller is designed based on model descretization or controller emulation, either a plant model or a controller discretization must be performed at each sampling time. Having such a discretized model computed analytically and pre-stored in the control system, saves a lot of computation power and memory in real-time. The same argument is true for computing approximate discrete-time models for nonlinear system \cite{abbaszadeh2008robust,abbaszadeh2016lipschitz}

\section{Software Packages}
Currently, to the best of the author's knowledge, there are two computer algebra systems that provide capabilities for control systems analysis and design, namely, \emph{Maple} and \emph{Mathematica}. Both Maple and Mathematica are major commercially available computer algebra systems widely used around the world. Here we briefly mention the coverage of the two systems with regards to dynamic system controls. A detailed comparison between the two systems support for controls is beyond the scope of this note. The interested reader is encouraged to consult the online documentation of each system.

The Maple computer algebra system, developed by \emph{Maplesoft}, provides a \emph{Dynamic Systems Package} containing functionalities for linear dynamic system representations (in transfer function, state space and pole-zero-gains, etc. formats), system interconnections, stability margins analysis and visualization, controllability and observability analysis, system transformations, linear systems simulations and continuous/discrete conversions, as well as linearization of nonlinear systems. The Maple Dynamic Systems Package is part of the standard Maple distribution and is fully integrated with the rest of the Maple computer algebra system. In addition, the \emph{MapleSim Control Design Toolbox}\footnote{Toolbox overview can be found at http://www.maplesoft.com/products/toolboxes/control\_design} utilizes Maple for symbolic and hybrid symbolic-numeric computations in control systems design. The toolbox provides full range of classical and modern control design techniques such as standard and advanced PID tuning methods, pole placement and observer design, LQR design and Kalman filter. The toolbox is an add-on to MapleSim, which a physical modeling software based on Maple and Modelica languages \cite{MapleSim}.

Similarly, the Mathematica computer algebra system, developed by \emph{Wolfram Research}, along with its \emph{SystemModeler}
\footnote{Toolbox overview can be found at http://www.wolfram.com/system-modeler}
Modelica-based physical modeling software, provides a suite of functionality for dynamical and control systems model representations, analysis and design of similar coverage \cite{SystemModeler}.

In addition to parameterized solutions as mentioned before, both systems provide remarkable features that are only possible through symbolic computations.
There are also many specialized and freely available code written by the research community using either Maple or Mathematica, each solving a particular problem in control analysis and design.


\bibliographystyle{IEEEtran}
\bibliography{References}

\end{document}